\documentclass[14pt,fullpage]{article}
\usepackage{calc}
\usepackage{lineno,hyperref}
\usepackage[a4paper,margin=1in,footskip=0.25in]{geometry}
\usepackage{setspace}
\usepackage{amsmath}
\usepackage{amsfonts}
\usepackage{relsize}
\usepackage[authoryear]{natbib}
\usepackage{setspace}
\usepackage{array}
\usepackage{graphicx}
\usepackage{hyperref}
\usepackage{geometry}
\usepackage{rotating}
\usepackage{afterpage}
\usepackage{subcaption}
\usepackage{verbatim}
\usepackage{setspace}
\usepackage{mathrsfs}

\begin{document}
	\onehalfspacing
	
	\begin{center}
		{\textbf{\large{Handling Missingness Value on Jointly Measured Time-Course and Time-to-event Data}}}
		
	\end{center}\vspace{1em}
	
	\begin{center}
		{\textit{Gajendra K. Vishwakarma $^a$, Atanu Bhattacharjee$^b$, Souvik Banerjee$^{a}\footnote[1]{Corresponding Author}$ }}		\vspace{1em}
		
			$^a$ Department of Mathematics \& Computing, 
		
		Indian Institute of Technology Dhanbad, Dhanbad-826004, India, 
		
		E-mail: vishwagk@rediffmail.com, souvik.stat@gmail.com
		
		$^b$ Section of Biostatistics, Centre for Cancer Epidemiology,
		
		Tata Memorial Center, Navi Mumbai, India 
		
		$^b$ Homi Bhabha National Institute, Mumbai, India

		E-mail : atanustat@gmail.com
		
	\end{center}

	\vspace{1em}
	
\begin{abstract}
		
	Joint modeling technique is a recent advancement in effectively analyzing the longitudinal history of patients with the occurrence of an event of interest attached to it. This procedure is successfully implemented in biomarker studies to examine parents with the occurrence of tumor. One of the typical problem that influences the necessary inference is the presence of missing values in the longitudinal responses as wel l as in covariates. The occurrence of missingness is very common due to the dropout of patients from the study. This article presents an effective and detailed way to handle the missing values in the covariates and response variable. This study discusses the effect of different multiple imputation techniques on the inferences of joint modeling implemented on imputed datasets. A simulation study is carried out to replicate the complex data structures and conveniently perform our analysis to show its efficacy in terms of parameter estimation. This analysis is further illustrated with the longitudinal and survival outcomes of biomarkers' study by assessing proper codes in R programming language.
		
	\end{abstract}

	
	\setlength{\parindent}{2em}
	\setstretch{1} 
	
	\section{Introduction}
	
	\onehalfspacing
Longitudinal measurements are receiving much attention in clinical research where patients are followed up to an event of interest. In these studies, the progression of disease can be closely monitored for a period of time and outcomes can be predicted more effcetively by assessing temporal changes in responses. The analysis of follow-up studies with time-to-event observations is called  Joint modeling of longitudinal and survival data (\cite{tsiatis1995, faucett1996})

This article is motivated by the gene expression data from Gene Expression Omnibus (GEO) database where longitudinal responses were collected for a number of liquid markers. The patients were given different treatments and the responses were collected at different time-points. It is examined whether any progression of tumor is developed for the patients during or after the treatment period. The problem with the dataset is there are several missing values in the longitudinal responses as well as covariates which affect the required inferences. The covariate missingness is an important aspect in this study as it differentiate the simple response missingness to a more complex missingness structure.  So, it is necessary to develop a proper statistical methodology to impute those missing values to gather additional information from the longitudinal and survival patterns. 

Gene expression data, collected from micro-array experiments, are useful to measure temporal changes in gene expression. These data play a crucial role in understanding the complex mechanism of genes and its effect on the disease progression. The tumor-specific gene expression profiling has an important contribution to biomarker detection (\cite{bhattacharjee2018}). The potentiality of early detection of disease or providing valuable information on disease progression presents the protein molecular biomarkers as a popular choice. Circulating biomarkers, which are obtained in biological liquid samples, have several advantages over tissue-based profiling due to their non-invasive nature and efficiency in longitudinal tracking of patients. However, the cost of collection and measurement of biomarkers is rising and hence the loss of significant amount of longitudinal responses is unaffordable and it is essential to retrieve the information lost based on the available data. 

During the study related to such a vulnerable disease, it often occurs that patients leave the study without any prior information (\cite{hui2013, bell2013}). As a result, this type of dataset contains missing observations due to drop-out of patients. Missingness can be divided into three different forms based on dependency pattern on the dataset, namely, Missing Completely at Random (MCAR), Missing at Random (MAR) and Missing not at Random (MNAR). In Bayesian and Likelihood methodology, MCAR and MAR mechanism are well addressed by introducing ignorable missingness and analyzing the observed data only (\cite{robins1995, hogan2004, tsiatis2007, bhattacharjee2019}). The MNAR mechanism comes under the non-ignorable category which requires more specifications of the joint distribution of the data and missing structure (\cite{ibrahim2009}). 

There are methods that simply ignore the covariates for which some observations are missing and analyze the remaining data as complete case analysis (\cite{littlerubin2014}). But in this process, the sample size is reduced and efficiency is lost. Also when the dataset is not large enough but the proportion of missing values is high, the estimates become biased (\cite{sunma2006, rubin1976}).  Some attempts have been made to address issues on covariates in joint modeling. Lu analyzed a joint modeling strategy for skew longitudinal survival data with missingness and mismeasured covariates using a logit model (\cite{lu2017}). A pseudo-likelihood approach was also implemented for longitudinal binary data with non-ignorable missing response and covariates (\cite{parzen2006}). Here a separate model has been considered for the missingness in covariates. Also, effort has been made in computing joint model with covariate subject to a limit of detection \cite{sattar2017}. Work has been performed for missing and left-censored time-varying covariates in joint modeling using some modified prior distributional assumption (\cite{chen2014}). However, there is a clear scarcity of proper modeling technique and imputation strategy for missing covariates in joint modeling context and it needs a detailed analytic and computational effort to show a pathway for the solution of this problem.

In joint models, a classical linear mixed effect model is popularly used to specify longitudinal measurements and Cox proportional hazard model for the survival part. The mixed effect model for longitudinal measurements is associated inside the survival model with an association parameter, (for example, see (\cite{wulfsohn1997, henderson2000, rizopoulos2012})). These characteristics enable joint modeling to capture the influence of covariates trajectory on the survival duration of patients. Bayesian and maximum likelihood techniques are efficiently utilized for parameter estimation (\cite{armero2016, song2016}). The joint modeling work was stimulated in problems depending on AIDS (\cite{tsiatis1995}) and currently it is becoming increasingly popular in other areas also (\cite{ibrahim2010}). 

There are several \textbf{R} software packages which serve as good tools to fit joint models for a continuous longitudinal data and a time-to-event process under maximum likelihood method (\cite{rizopoulos2010}). Bayesian approach has also been employed in several packages to estimate the parameters and provide more accurate results in many instances (\cite{rizopoulos2016}). Some recent packages use Monte Carlo Expectation-Maximization algorithm for estimation of parameters and consider multivariate Gaussian process for joint modeling (\cite{philipson2018, hickey2018}). The \textbf{mice} (Multivariate Imputation of Chained Equations) package in \textbf{R} provides a good solution for the complex incomplete data problems (\cite{buuren2010}). It generates the missing observations using conditional densities by MCMC procedure and specifies different imputation models for different covariates. Some other packages use hot-deck imputation, k-nearest neighbor imputation, regression imputation etc. Other successful efforts considered expectation-maximization with bootstrapping (EMB) algorithm to impute missing observations considering the normality assumption and MAR (missing at random) missingness structure (\cite{honaker2011}). Iterative robust model-based imputation approach, implemented through \textit{irmi()} function in \textbf{VIM} (\cite{kowarik2016}) package, also provides a good alternative to impute missing observations using robust methods (\cite{templ2011}). We use \textbf{R} open-source software for the simulation of posterior samples for the parameters using Markov Chain Monte Carlo (MCMC) method.

In section 2, we describe the estimation framework of joint models for longitudinal and time-to-event data and also explain the imputation strategies for missing data with covariate missingness. In section 3, a simulation study is described to explain the mentioned methodology. In section 4, we perform our analysis on a real dataset of gene expression to show its efficacy. In section 5 and 6, we discuss some results on the proposed technique and also add future scopes of this methodology.
	
	\section{Methodology}
	\subsection{Joint Modeling}
	
	In this section, we have discussed the model-building techniques to analyze the longitudinal and survival measurements and the joint modeling method to obtain the efficient treatment strategy. 
	
	Longitudinal data is a collection of repeated observations of a particular subject over a period of time (\cite{weiss2005}). The temporal ordering of longitudinal data is important as measurements that are close to each other tend to similar than the measurements collected at a distant time-point. So, the analysis also gets difficult with complex variance structure.    
	
	One of the widely used tools for analysis of longitudinal measurements is mixed effect model. The mixed effect model is defined as, 
	
	\begin{equation}
	\begin{split}
	Y_i(t)=M_i(t)+\epsilon_i(t)=X_i(t)\beta+Z_i(t)b_i+\epsilon_i(t),\\
	~~~b_i\sim N(0,D),~~\epsilon_i(t)\sim N(0,\sigma^2)
	\end{split}
	\end{equation}
	
	where $Y_i(t)$ denotes the longitudinal measurement of $i^{th}$ patient at time $t$, $M_i(t)$ is the true effect of the data or, the true longitudinal response which we will model in mixed effect model, $\epsilon_i(t)$ is the error component attached to the data. $X_i(t)$ \& $Z_i(t)$ are the fixed effects and random effects respectively. Also, $\beta$ \& $b_i$ are the coefficients of fixed effect and random effects respectively.  We assume the measurement error component $\epsilon_i(t)$ follows a Normal distribution with mean 0 and variance $\sigma^2$ and the error is independent of $b_i$. The estimates of the coefficients can be computed using generalized estimating equations for unknown correlation structures between outcomes.
	
	Dataset consisting of time to the event of interest is referred to as time-to-event data. When the event of interest is defined as occurrence of death or any disease to the patient enlisted in the subject, then this kind of dataset is referred to as Survival Data. Observations are called censored when the event of interest does not appear for the patient till the time he is available in the study. Generally, this kind of data is continuous in nature. 
	
	Non-parametric methods like Kaplan-Meier estimator is widely used to estimate and plot the survival probabilities as a function of time. It can also be used for comparison of multiple groups of subjects. For survival data, often the interest lies in the association of hazard function and related covariates. A popularly used parametric model for survival data is Cox proportional hazard model which is defined as,
	
	\begin{equation}
	h_i(t|K_i)=h_0(t)exp(K_i\gamma)
	\end{equation}
	
	where $h_i(t|K_i)$ is the hazard rate for the $i^{th}$ patient who experiences the event of interest at time t, $h_0(t)$ is the baseline hazard function and $K_i$ is baseline covariates with corresponding regression coefficient $\gamma$. Some popular choices of baseline hazard function are piecewise constant function, Weibull risk function and unspecified risk function. In most applications, one of the main objectives is to estimate the $\gamma$ which explains how the covariates are related to the hazard function.  
	
	In joint modeling, we associate both the mixed effect model of longitudinal measurements and Cox model of survival outcomes in a single model and estimate the coefficients (\cite{wulfsohn1997, tsiatis2001}). The joint model is explained as, 
	\begin{equation}
	\begin{split}
	h_i(t|K_i,M_i)~~~~~~~~~~~~~~~~~~~~~~~~~~~~~~~~~~~~~~~~~~~~~~~~~~~~~~\\
	=\lim_{dt\rightarrow 0} Pr(t\leq T_i^*<t+dt|T_i^*\geq t, K_i,M_{ij})/dt
	\end{split}
	\end{equation}
	
	\begin{equation}	
	\hspace{-3em}=h_0(t)exp(K_i\gamma+M_{ij}\alpha)~~~~~~~~~~~~~~~~~~~~~~~
	\end{equation}
	
	Here $\alpha$ measures the effect of longitudinal process on the time-to-event measurement. It serves as a link between the risk function of cancer infection at time $t$ with the underlying longitudinal model. The other parameters have their usual meanings as explained in the longitudinal and survival process. One of the advantages of this joint model lies in the fact that the parameters are estimated simultaneously in this approach. 		
	
	The log-likelihood of the joint model as described by Rizopoulous (\cite{rizopoulos2012}) is as follows,
	
	\begin{equation}
	log~ p(T_i,\delta_i,y_i; \theta)=log \int p(T_i,\delta_i,y_i,b_i; \theta) db_i ~~~~~~~~~~~~~~~~~~~~~~~~~~~~~~~~~~
	\end{equation}
	
	\begin{equation}
	=log \int p(T_i,\delta_i|b_i; \theta_t)[\prod_{j}p\{y_i|b_i;\theta_y\}]p(b_i;\theta_b)db_i
	\end{equation}
	
	$\delta_i$ is the censoring indicator where 1 implies censored observation and 0 implies uncensored observation. $\theta=(\theta_b,\theta_t,\theta_y)$ denotes full parameter vector where $\theta_t$ is parameters for time-to-event data, $\theta_y$ denotes the parameters of longitudinal observations and $\theta_b$ denotes the parameters for random effect component. The parameters of the joint model are estimated using the maximum likelihood method and the random effects are predicted using their conditional expectations given all of the data. Interpretations of the parameters remain the same as what is explained in random effect model and survival model. Direct maximization of this model is impossible to work on due to its complex nature and non-parametric components. The expectation-maximization(EM) algorithm is one of the most used techniques for optimization purposes of joint likelihood (\cite{wulfsohn1997, henderson2000}).
	
	In bayesian framework, the parameters of the model are estimated using Markov Chain Monte Carlo(MCMC) method. The posterior distribution of the joint model is defined as,
	
	\begin{equation}
	p(\theta,b_i|T_i,\delta_i,y_i) \propto p(T_i,\delta_i|b_i,\theta)p(y_i|b_i,\theta)p(b_i|\theta)p(b_i|\theta)p(\theta)
	\end{equation} 
	where $\theta$ denotes the vector of all the parameters. So, survival likelihood for patient $i$ is, 
	\begin{multline}
	p(T_i,\delta_i|b_i,\beta,\theta_s)=h_i\{T_i|M_i(T_i),\theta_s\}^{\delta_i}S_i\{T_i|M_i(T_i), \theta_s\}\\
	=[h_0(T_i|\gamma_s)exp\{\gamma^T K_i+\alpha M_i(T_i)\}]^{\delta_i} \times exp \Big(-\int_{0}^{T_i} h_0(s|\gamma_s)exp\{\gamma^T K_i+\alpha M_i(s)\}ds\Big)
	\end{multline}
	Here $\theta_s=(\gamma_s,\gamma,\alpha)$ where $\gamma_s$ denotes the parameters associated with the B-splines for the baseline hazard. An inverse Wishart prior is considered for the variance-covariance matrix ($D$) and a gamma prior is used for the variance of errors ($\sigma^2$). 
	
    The Bayesian approach offers the use of Markov Chain Monte Carlo (MCMC) sampling algorithm to estimate the posterior distribution of the parameters. The joint models are analyzed using R programming software. 
	
	\subsection{Missing Data Methodology}
	
	In real-life situations, obtaining a dataset without any missing observations especially for patients with diseases is a difficult job. A subject can refrain itself to be measured at follow-up times which creates a monotone missing data pattern in longitudinal study. Also, there are some instances when the subject is absent at one follow-up time and then present again at subsequent time-points, resulting in intermittent missingness. It is difficult to evaluate the likelihood functions with non-monotone missing data patterns as simple factorization rule cannot be applied here. However, in cases where missing data pattern is considered to be missing at random (MAR), some imputation models using the software can be applied to estimate the missing data. Similarly, in survival studies, datasets regarding the death of a patient often become unavailable generating censored observations. The covariates or factors impacting upon the failure time are also not available always throughout the entire study. The difference of measuring frequencies and the inability to measure all the covariates at a certain time slot will also result in incomplete data.      
	
	A very efficient way to analyze dataset consists of missing observations is considering the ignorability condition. This condition comes into the picture when the missingness pattern is MAR and full-data parameters can be decomposed into two independent parts: one signifies the full data response model and the other exhibits missing data mechanism. The ignorability condition does not interpret to drop the missing observations, rather implies that the missingness mechanism can be left unspecified to draw posterior inference of the full data model. When the ignorability condition is believed to be not a suitable option, we can use a more general approach where missingness depends on the unobserved part of responses given the observed part and covariates. 
	
	In MAR, the missingness depends only on the observed values, not on the unobserved ones (\cite{templ2011}). In our study, it is found that the patients tend to skip the clinical tests due to health reasons whenever their condition is worsening or the size of the tumor is big enough. Also, it can be found that some of the patients do not feel obliged to test often when values are not much significant and importance of further testing is not felt. So, the missingness depends on the observed value of the biomarker i.e., the missingness is Missing at Random (MAR).
	
	Four different mechanisms that have been used in this article for multiple imputations are namely, Expectation-Maximization Bootstrap Algorithm, Predictive Mean Matching, Classification \& Regression Tree and Bayesian Linear Regression Method. The multiple imputation algorithms with covariate missingness are discussed here. The details on these methods are explained in the following works (\cite{buuren2010, honaker2011, rubin2004, doove2014,vishwakarma2016}).
	
	\paragraph{Classification and Regression Tree (CART)}
	
	Cart method is a popular machine learning algorithm which identifies predictors and cutpoints in the predictors that are used to split the dataset into more homogeneous subsamples. This process is repeated to obtain a series of splits that define a binary tree. The imputation algorithm in CART method is,
	
	(i)Draw a bootstrap sample of $(\dot{y}_{obs},\dot{X}_{obs})$ of size $n_1$ from $(y_{obs},X_{obs})$
	
	(ii) Fit $\dot{y}_{obs}$ by $\dot{X}_{obs}$ by a tree model $f(X)$.
	
	(iii) Predict $n_0$ terminal nodes from $f(X_{mis})$.
	
	(iv) For $y_{mis}$, determine in which leaf they will end up according to the tree formed in setp-(iii).
	
	(v) Randomly select one donor $i_j$ of the leaf ended up in step (iv) and replace the  missing value $\dot{y}_j$ by $y_{j}$, $j=1,2,...,n_0$
	
	(vi) Repeat the steps (ii) to (v) several numbers of times. 
	
	(vii) The same procedure is performed for each variable in the dataset replacing the variable containing missing observations with the other variables for which the data available. Thus at each iteration, we obtain imputed values for each variable.
	
	\paragraph{Expectation-Maximization Bootstrap (EMB) Algorithm}
	
	The EMB algorithm is an efficient method for multiple imputation technique. Here the objective is to break the distribution of observed data and missingness into other parts where the missingness likelihood can be explained by observed data. Under MAR conditions, let $D^{obs}$ denotes the observed data and $M$ the missingness matrix. So, we get,
	
	\begin{equation}
	p(D^{obs},M|\theta)=p(M|D^{obs})p(D^{obs}|\theta)
	\end{equation}
	where $\theta$ is the complete data parameters. In this imputation technique, it is assumed that the complete data follow multivariate normal distribution i.e., $D \sim N_k(\mu,\Sigma)$. Now, the parameters will be estimated as
	\begin{equation}
	L(\theta|D^{obs}) \propto p(D^{obs}|\theta) =\int p(D|\theta) dD^{mis}
	\end{equation}
	
    The data is first divided in $m$ bootstrapped samples. Then for each sample, the EM algorithm is performed to produce point estimates of $\mu$ and $\Sigma$. Thus we impute the missing observations using those estimates and original sample units. Analysis is performed for each of the samples and in the end, the results are combined by averaging the estimated values.   

\paragraph{Predictive Mean Matching}

In this method, the following algorithm is maintained to impute the missing observations.

(i) For cases with no missing data, a linear regression of $y$ on $X$ is fitted and estimates of the coefficients are obtained.

(ii) A random draw is made from the posterior distribution of coefficients. Typically this can be a random draw from a multivariate normal distribution.

(iii) Using those random draws of the coefficients, the predicted values of $y$ are generated.

(iv) For each missing $y$, a set of cases with observed $y$ whose predicted values are close to the observed ones according to any metric defined is chosen.

(v) From those cases, we randomly choose one and assign its  observed value to the missing observation.

(vi) Repeat the process to generate multiple datasets.

This purpose of this method is to match cases with missing data to similar cases found in the observed dataset.

\paragraph{Bayesian Linear Regression Method} 
	
	In this method, the missing observations are imputed using Bayesian linear regression model. The imputation algorithm is, 
	
	(i) Calculate $V=(S+diag(S)\kappa)^{-1}$ with small $\kappa$ where $S=X_{obs}'X_{obs}$
	
	(ii) Calculate the regression weights $\hat{\beta}=VX_{obs}'y_{obs}$
	
	(iii) Calculate $\sigma^2=(y_{obs}-X_{obs}\hat{\beta})'(y_{obs}-X_{obs}\hat{\beta})/g$ where $g$ is a random variable drawn from $\chi^2_(n-q)$
	
	(iv) Calculate $\dot{\beta}=\hat{\beta}+\dot{\sigma}\dot{z}V^{1/2}$ where $V^{1/2}$ is computed using Cholesky decomposition and $\dot{z}$ are $q$ random draws from $N(0,1)$
	
	(v) Calculate $\dot{(y)}=X_{mis}\dot{\beta}+\dot{z_1}\dot{\sigma}$ where $\dot{z_1}$ is $n_0$ independent variates from $N(0,1)$

	\section{Simulation Study}
	
	\subsection{Design}
	A simulation study was performed to find out the effect of missing observations on the inference and estimates of parameters. First, we simulated a full dataset using fixed estimates of the parameters for both longitudinal and survival model and also selected an association parameter for the joint model. The submodel for the longitudinal outcome is defined as,
	\begin{multline}
	y_i(t)=\beta_1+\beta_2 t+\beta_3*(t\_stage)_{1i}+\beta_4*(t\_stage)_{2i}+\beta_5*(t\_stage)_{3i}+\beta_6* (n\_stage)_{1i} +\\
	b_{i}+\epsilon_i(t)  
	\end{multline}

	where $\beta$'s define the fixed effect parameters and $b$ defines the subject specific variations which are assumed to be normally distributed $N(0,\sigma^2)$. The variance-covariance matrix $D$ ($=\sigma^2$, as $b$ is univariate. D is used to keep consistency with the notation) of the random effects is left unstructured. The $t\_stage$ and $n\_stage$ are two categorical covariates; $t\_stage$ has 4 levels namely 0,1,2,3 and $n\_stage$ has 2 levels namely 0 and 1. The levels are ordinal in nature and it is assumed that higher the value, greater the effect of the covariate. The levels of $t\_stage$ and $n\_stage$ are drawn using equal probability for all levels. While choosing the time component, we kept in mind the treatment plan for patients in the actual scenario. There are 4 patient visit times or sample collection times considered for simulation of longitudinal outcomes and the visit times are chosen from the following distributions: $Poisson(0.7),N(30,5), N(60,5)$ and $N(90,5)$ respectively. 
	
	The survival outcomes and times are generated by,
	
	\begin{equation}
	h_i(t)=h_0(t)exp(\gamma_1*(t\_stage)_{1i}+\gamma_2*(t\_stage)_{2i}+\gamma_3*(t\_stage)_{3i}+\alpha m_i(t))
	\end{equation}
	
	$m_i(t)$ is considered as the true longitudinal response at time $t$. The starting parameters are drwan randomly from different distributions to maintain the independence of the regression coefficients and incorporate the difference in estimation procedure under different methods. The distributions considered for different parameters are as follows: $\beta_1 \sim U(0,10), \beta_2 \sim U(0.5,1), \beta_3 \sim U(0.5,1), \beta_4 \sim U(0.7,1.2), \beta_5 \sim U(0.8,1.5), \beta_6 \sim U(0.5,1.2), \gamma_1 \sim U(-0.5,0.5), \gamma_2 \sim U(-1,1), \gamma_3 \sim U(-1,1), \sigma \sim U(0.5,2)$. The baseline hazard function is generated using Weibull distribution with shape parameter 5 and the association parameter $\alpha$ is generated using $U(0.5,1.5)$. $D$ or, Var(b) is generated using $U(0.5,2)$. The censoring time is simulated from an Exponential distribution with mean 2. To include MAR missingness in the simulation study, the biomarker values are missing with higher probability for higher the values of $t\_stage$ and $n\_stage$. However, the missingness in covariates is randomly generated. The covariate missingness is considered to be around 20\% of the total observations.
	
	\subsection{Analysis}
	
	Multiple Imputation (MI) technique is employed to generate the missing data. Here we use 4 different multiple imputation methods such as Expectation-Maximization Bootstrap Algorithm, Predictive Mean Matching, Classification \& Regression Tree and Bayesian Linear Regression Method. Imputed datasets are obtained by using of MICE and Amelia package in R open-source software. For each of the MI techniques, we obtained 4 imputed datasets and analysis was carried out on each  dataset. Once the entire analysis is carried out, the results are combined to obtain the estimates of the parameters and they are compared with the true parameter values considered for simulation. 
	
	We genearted 5000 sets of starting values of the paramters and repeated the simulation study and corresponding analysis. For each generated dataset containing missing observations, we employed the mentioned missing data techniques and obtained 200 imputed datasets for each of them to incorporate the sampling fluctuations among the imputed samples. The joint modeling analysis is performed on each of the those imputed datasets and results are combined. Thus, as a whole, the simulation study is replicated 5000 times with 200 imputed samples for each generated data for each multiple imputation method.
	
	A linear mixed effect model is considered for the longitudinal process with $t\_stage$ and $n\_stage$ are considered as covariates. Proportional hazard model is considered for the survival outcome with the baseline hazard function is assumed to follow Weibull distribution. We considered two different approaches for estimating the parameters of joint models namely maximum likelihood approach and  bayesian approach. Results for both the methods are compared and the estimated parameters are checked with the original values. In ML approach of joint modeling, we considered unspecified model for baseline hazard function whereas for bayesian approach, the baseline risk function is approximated using splines. Gauss-Hermite quadrature integration was used for the estimation of parameters in ML approach and Markov Chain Monte Carlo was used for parameter estimate in bayesian procedure.
	
	The effect of missing observations in response and covariates has been assessed using two different joint modeling approach. The difference in parameter values in the simulation study has been examined and on the basis of the estimates, the multiple imputation technique are judged. The estimates of the parameters are obtained for each generated dataset and the parameter estimates (for ML methods) or posterior means (for Bayesian method) and standard errors (for ML method) or standard deviations (for Bayesian method) are averaged. As the starting parameters are randomly generated from different distributions, we compared the performance of different imputation techniques through two performance measures namely Coverage Probability and Root Mean Squared Error (RMSE). The performance measures of the parameters for both response and covariate missingness is displayed in Tables \ref{table:simcov_bayes_data} \& \ref{table:simcov_ml_data} and the same for only response missingness are presented in Tables \ref{table:simwb_bayes_data} \& \ref{table:simwb_ml_data}. 
	
	Among different multiple imputation techniques, the data imputed by Predictive Mean Matching (PMM) method results in highest coverage and lowest RMSE among the four methods followed by EMB algorithm. The Bayesian Normal Regression (NORM) and Classification and Regression Tree (CART) method performs close to each other while CART methods performs poorest among those four methods. The coverages are higher for joint modeling analysis using maximum likelihood method than bayesian method. Also, the models show better performance when the missingness is occurred in response only. The RMSEs are much lower for the association parameter $\alpha$ resulting in better estimate. Among different imputation methods, most of the estimates obtained from imputed data using EMB algorithm are found to be close to the full data analysis. For response missingness only, all the four methods perform better and they are close to each other while PMM method also performs slightly better than the others.

	\section{Analysis on Gene Expression Study}
	
	 To perform our analysis, we have used a published dataset on Gene Expression study collected from Gene Expression Omnibus (GEO) database (\url{https://www.ncbi.nlm.nih.gov.in/geo/}) under the accession number GSE65622. In this study, all the patients were given neoasjuvant chemotherapy (NACT) followed by chemoradiotherapy (CRT) and treatment was evaluated for four weeks after CRT completion. After 2-4 weeks, radical pelvic surgery was considered. The study consists dataset on 80 patients and the observations were obtained in the following manner. Serum samples were collected at baseline (Sapling point=1), following NACT post-NACT(Sampling point=2), at CRT completion (Sampling point=3), post-CRT (Sampling point=4) and at treatment evaluation (Sampling point=5). With a high-density antibody array (AHH-BLG-1;RayBiotech Inc.), the samples were analyzed for the presence of 507 proteins. Further explanation about the study setting and detailed description of the data can be cited from Meltzer et al. (2016). The objective of the study is to classify the proteins according to their effect on cancer progression and investigate the association between factors regarding tumor severity and duration of disease reoccurrence.
	 
\begin{figure}[!h]
		\centering
		\includegraphics[width=\linewidth]{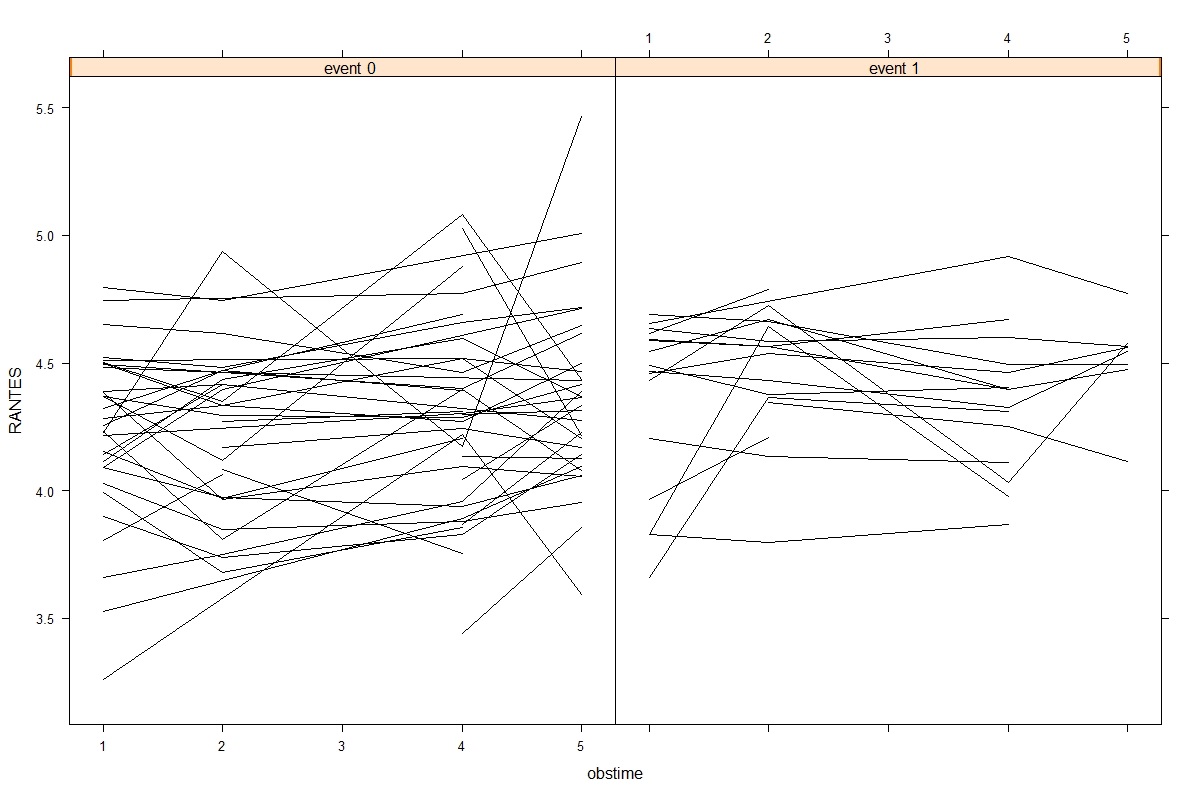} 
		\caption{Profile plot of log values of the Rantes Biomarker for different patients }
		\label{fig:profile_rantes}
	\end{figure}

\begin{figure}[!h]
		\centering
		\begin{subfigure}[b]{0.43\linewidth}
			\includegraphics[width=\linewidth]{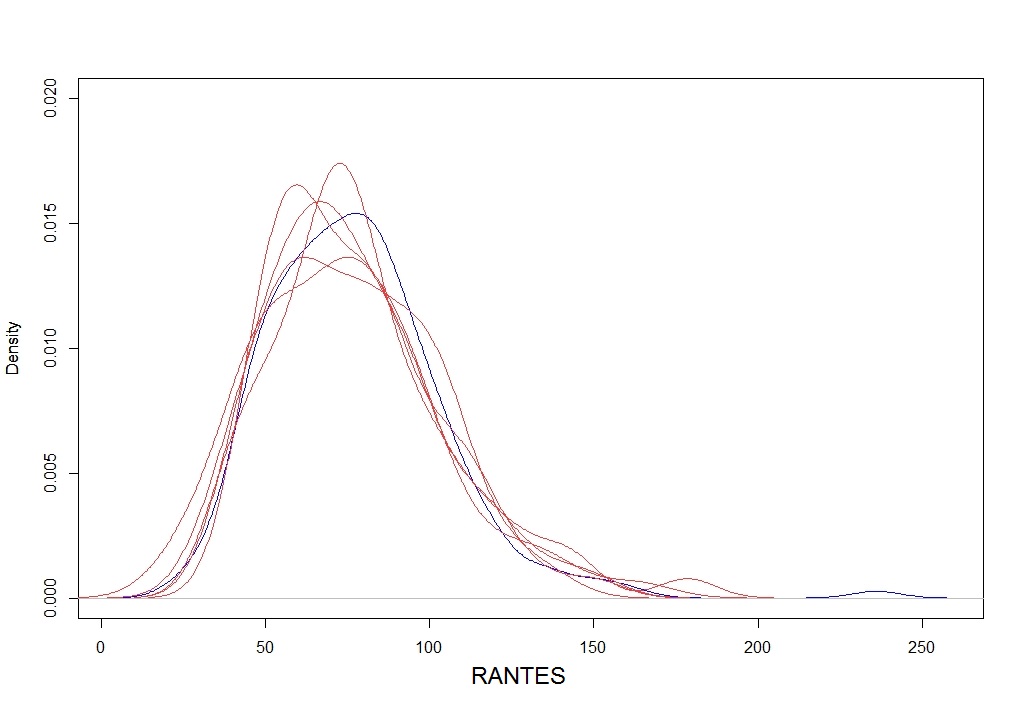}
			\caption{}
		\end{subfigure}
		\begin{subfigure}[b]{0.45\linewidth}
			\includegraphics[width=\linewidth]{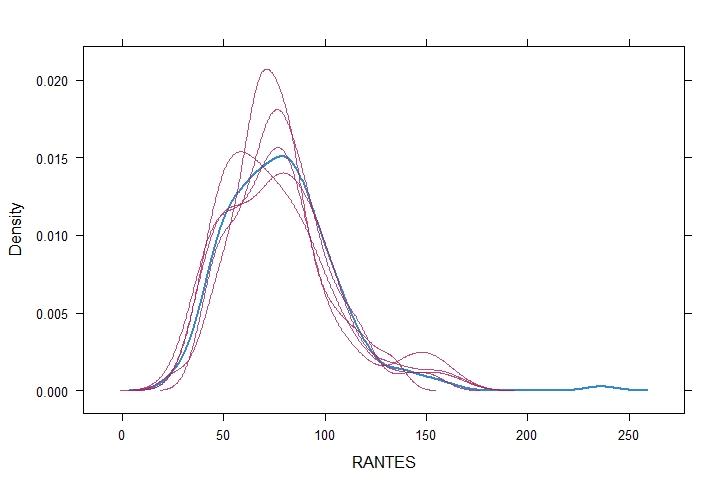}
			\caption{}
		\end{subfigure}
		\begin{subfigure}[b]{0.45\linewidth}
			\includegraphics[width=\linewidth]{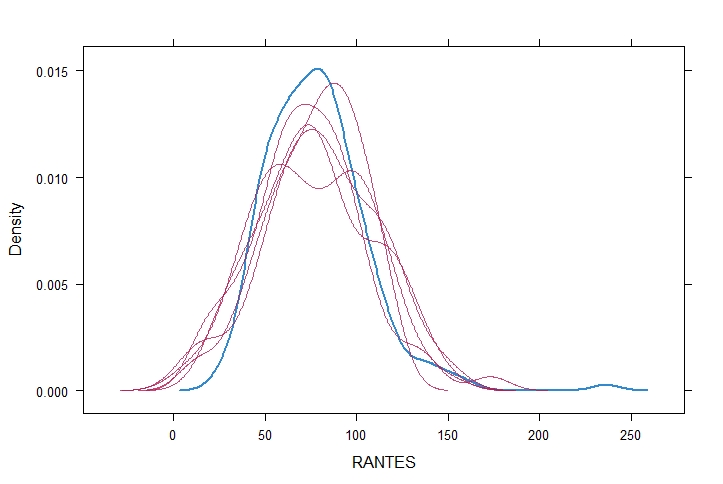}
			\caption{}
		\end{subfigure}
		\begin{subfigure}[b]{0.46\linewidth}
			\includegraphics[width=\linewidth]{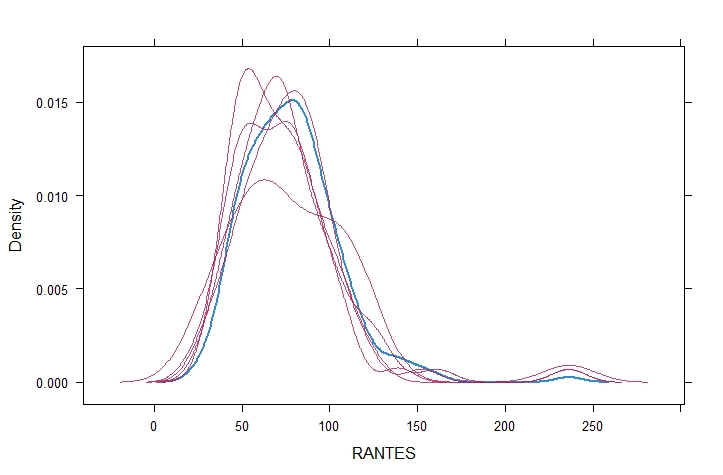}
			\caption{}
		\end{subfigure}
		\caption{Density comparisons of imputed data for 4 different imputation techniques applied on Biomarker data (a) EMB Algorithm, (b) CART method, (c) NORM method, (d) PMM Method. 5 imputed datasets per method were generated and plotted with the observed biomarker values for comparison.}
		\label{fig:imputation_density}
	\end{figure}

	 In the biomarker study, measurements on several biomarkers are gathered from the patients. It is our interest to find how the missing observations in $Rantes$, a well-established biomarker related to pulmonary tuberculosis and HIV disease, affects the analysis of joint modeling. The $Rantes$ biomarker is also known as $CCL5$ and regulated upon activation, normal T-cell expressed and secreted. It plays a crucial role in the stimulation of T-cell proliferation and activates the release of proteins in anti-mycobacterial immunity \cite{bacon1995}. The distribution of Rantes is positively skewed, so a log transformation is applied to the response.The transformed data is ranging from 3.241 to 5.465 with mean 4.348 and standard deviation 0.326. 65 observations were missing out of 285 records i.e., around 22.81\% observations were missing in the response which consists of 38 patients. Further analysis shows that the missingness has occurred for 15 out of 28 patients who experienced the event whereas the responses of 23 out of 51 patients who did not experience the event are missing. The information regarding the trg\_score are also missing for 3 patients which encourages our study of covariate missingness in biomarker data. For most of the patients with missing responses have high t-stage and n-stage values. Hence, we assume that the missingness  structure is dependent on the observed data (MAR missingness). In follow-up studies of liquid biopsies, patients generally provide responses at early stages which lead to low number of missing data at the early stages. However, when they do not find significant improvements in the biomarker values, they tend to leave the study which leads to missingness. The trajectories of longitudinal response in shown in the profile plot displayed in Figure \ref{fig:profile_rantes}. Four different multiple imputation algorithms, as described before, are used for imputing datasets. 5 datasets are generated for each of the imputation methods with necessary iteration and the joint modeling is applied on the imputed dataset. The density comparisons of the imputed data with original data are shown in Figure \ref{fig:imputation_density}.
	
	Joint modeling approach was followed to which we explicitly used a linear mixed effect model for the longitudinal values of the logarithm of biomarker values. In this study, we considered the longitudinal responses till the occurrence of event or censoring time, whichever is earlier and rest of the responses are ignored. In particular, due to randomization technique used in the study, we considered the fixed effects part of the longitudinal sub-model as the effect of sampling time (expressed in no. of days and scaled in [0,1] by dividing with 365), t-stage, n-stage and trg-score. The scores are independent and hence no interaction term among the covariates is considered. In the random effects part, we consider a patient-specific random intercept model.  
	
	The longitudinal submodel considered for the gene expression data is,
	
	 \begin{multline}
	 log(y_i(t))= \beta_1+\beta_2*(days/365)_i+\beta_3*t\_stage_i+\beta_4*n\_stage_i+
	 \beta_5*trg\_score_i+b_i+\epsilon_i(t)
	 \end{multline}
	 where t\_stage, n\_stage and trg\_score are the ordinal covariates where higher value denotes more severe effect.
	 
	In the survival model, we used Cox Proportional Hazard Model using t-stage and n-stage as covariates. The survival submodel is defined as,
	
	\begin{equation}
	h_i(t)=h_0(t)exp(\gamma_1*t\_stage_3+\gamma_2*t\_stage_4++\gamma_3*n\_stage_1+\gamma_4*n\_stage_2+\alpha m_i(t))
	\end{equation}
	The baseline hazard maximum likelihood approach is left unstructured. In bayesian approach, it is modeled with cubic B-splines with 5 knots are being on the percentiles of the observed event times. 
	
	Both bayesian and ML (maximum likelihood) framework has been used for joint modeling. In ML framework, the baseline risk function is left unspecified whereas in bayesian framework, the baseline risk function is approximated using splines. Also in bayesian framework, a simple association structure has been used where the longitudinal model is associated with the survival covariates with an association parameter. We have considered only the observations which have occurred before the event. The estimates of the parameters are obtained through joint modeling for each generated dataset and the parameter estimates and standard errors (for ML estimate) or standard deviations (for bayesian method) are averaged.  Comparison for the bayesian models and ML models was done using Deviance information criterion (DIC) \& Akaike Information Criterion (AIC) values respectively and they are shown in Tables \ref{table:real_ml_estimate} and \ref{table:real_bayes_estimate}. 
	
	From the tables, we can find that all the imputation models are pretty close to each other and estimates of some of the parameters are different in complete case analysis than imputed data analysis. In complete case analysis, we choose only those patients where all the longitudinal outcomes till the time of event are available. By comparing the AIC values, it can be said that the PMM, CART and EMB algorithm imputation methods are very close to each other whereas NORM method has higher AIC showing a less preferable method. The association parameter between longitudinal biomarker and survival event obtained through ML estimate is nearly 1.65 for all the cases except EMB algorithm for which the estimate is 2.015. However, for bayesian method, it is approximately 1.10 for all the methods except EMB algorithm for which the estimate is 1.916.

	\section{Discussion}
	Joint modeling of longitudinal and survival data with missing covariates is a challenging area in oncology research. The presence of missing observations makes it difficult to analyze the data and often leads to biased inference. Available techniques based on complete observations often leads to biased estimate when the proportion of missingness is high enough in the dataset. The joint models combine the conventional linear mixed effect model for longitudinal responses with the renowned Cox proportional hazard model for the survival process. Maximizing the likelihood is not always possible in straightway due to the non-explicit form of the likelihood function which leads to the use of numerical techniques and bayesian methodology.
	
	It is more challenging to compute the conditional distribution of the missingness mechanism present in the joint models. Listwise deletion method to get the complete data is valid only under MCAR conditions, but not for MAR and MNAR situations. More specifically, when the probability of missingness is dependent on the unobserved responses and covariates, it requires special techniques to handle such scenarios.
	
	Separate modeling mechanism is necessary to postulate the cases to obtain valid inference for joint models. Three main modeling frameworks for missing observations are a pattern mixture model, shared parameter model, and selection model. However, these models do not take into account the missingness of the covariates and are mainly concerned with the response variable.
	In our study, we used multiple imputation techniques to obtain the missing observations in the covariates also. Among several algorithms, we took support of four methods for data imputation through EMB algorithm, Correlation and regression tree, Bayesian normal regression, and Predictive mean matching. The Bayesian perspective in the form of EMB algorithm for imputation has produced better estimates among those methods. Prior distributional assumptions are used to generate statistical inferences from the posterior distributions of the unknown variables including the missing data. Well-known packages in \textbf{R} programming software were used to analyze the dataset.
	
	\section{Conclusion}
	
	In this article, we demonstrated an effective way to analyze data in joint modeling scenario when the response and covariates both have missing observations. Missing covariates for joint longitudinal and survival data modeling is a challenging task to solve. Our approach shows an efficient way to handle such missingness in Biomarker study with different multiple imputation techniques and their effects on joint modeling. Both the joint modeling approach is assessed to extensively examine the effect of missing observations on different estimation procedures. In Biomarker study, the cost of obtaining dataset is high, so it is necessary to use proper techniques to handle missing observations.
	
	Our procedure is still limited in handling missingness in the parametric approach. However, further research work is required to accommodate the non-gaussian response process. The semi-parametric approach could be an attractive alternative choice for multiple imputations. It is more flexible than others with extension for non-Gaussian response structure to fit joint models.
	
	\section{Supplementary Material}
	
	The detailed R programming code used in this article is available in the following link \url{https://github.com/souvikbanerjee91/missing-data-in-joint-modeling.git} 
	
	\section{Acknowledgment}
	
	Authors are deeply indebted to the editor  Prof. Narayanaswamy Balakrishnan and learned referee for their valuable suggestions leading to improving the quality of contents and presentation of the original manuscript. Authors are also thankful to Science and Engineering Research Board, Department of Science \& Technology, Government of India, for providing necessary support to carry out the present research work through project Grant No. EMR/2016/003305.

	\newpage
	
	\newpage

	\newgeometry{top=2in}
	\afterpage{\restoregeometry}
	
\centering

\begin{table}[!h]
	\caption{Coverage Probability and Root Mean Squared Error of Bayesian Estimates of Coefficients in Joint Models on Simulated Biomarker Data with both response and covariate missingness (The RMSEs are given in the brackets)}
	\begin{center}
	\begin{tabular}{|l|l|l|l|l|l|}
		\hline
		Variable & \begin{tabular}[c]{@{}l@{}}Full \\ Dataset\end{tabular} & \begin{tabular}[c]{@{}l@{}}EMB\\ Algorithm\end{tabular} & \begin{tabular}[c]{@{}l@{}}PMM\\ Method\end{tabular} & \begin{tabular}[c]{@{}l@{}}NORM\\ Method\end{tabular} & \begin{tabular}[c]{@{}l@{}}CART \\ Method\end{tabular} \\
		\hline
		$\beta_1$  & 0.97  (0.13)                                             &
		 0.81   (0.22)                                                &
		 0.91   (0.18)                                             & 
		 0.76    (0.24)                                             & 
		 0.72    (0.26)                                              \\
		$\beta_2$  & 0.98  (0.21)                                                 & 
	    0.74     (0.29)                                             & 
		0.82       (0.23)                                         & 
		0.71       (0.28)                                          & 
		0.67        (0.32)                                          \\
		$\beta_3$  & 0.95   (0.13)                                                & 
		0.74      (0.23)                                            & 
		0.86      (0.22)                                         & 
		0.70       (0.28)                                          & 
		0.64       (0.39)                                          \\
		$\beta_4$  & 0.96   (0.17)                             &
		  0.73   (0.25)                                               & 
		0.84      (0.18)                                         & 
		0.68       (0.30)                                         & 
		0.61        (0.28)                                          \\
		$\beta_5$  & 0.97    (0.22)                                  &
		  0.68  (0.23)                                        & 
		  0.83   (0.26)                                             & 
		  0.72    (0.31)                                             &
		  0.66     (0.34)                                             \\
		$\beta_6$  & 0.98   (0.11)                                               & 
		0.78   (0.19)                                              & 
		0.81     (0.13)                                          & 
		0.65     (0.22)                                           & 
		0.62      (0.24)                                           \\
		\hline
		$\gamma_1$ & 0.97    (0.22)                                & 
		 0.67  (0.31)                                            & 
		 0.78   (0.29)                                            & 
		 0.67   (0.32)                                             & 
		 0.60   (0.33)                                                \\
		$\gamma_2$ & 0.98    (0.24)                              &
		 0.67 (0.35)                                                 & 
		0.77   (0.29)                                            & 
		0.72    (0.37)                                            & 
		0.62     (0.35)                                          \\
		$\gamma_3$ & 0.96   (0.23)                              &
		  0.72  (0.35)                                                & 
		  0.83   (0.28)                                            & 
		  0.73   (0.34)                                             &
		  0.71   (0.37)                                              \\
		\hline
		$\alpha$    & 0.98   (0.15)                              &
		 0.71   (0.24)                                                &
		  0.84    (0.22)                                            & 
		  0.66    (0.28)                                             & 
		  0.68     (0.31)                                             \\
		D        & 0.97  (0.06)                                    &
		 0.73   (0.20)                                               & 
		 0.75  (0.15)                                          & 
		 0.69    (0.26)                                             & 
		 0.62    (0.28)                                              \\
		$\sigma$        & 0.96  (0.11)                                   &
		0.68   (0.19)                                                & 
		0.72   (0.17)                                          & 
		0.67    (0.23)                                             & 
		0.61    (0.26)                                              \\ 
		\hline
	\end{tabular}
\end{center}
\label{table:simcov_bayes_data}
\end{table}

\begin{table}[!h]
	\caption{Coverage Probability and Root Mean Squared Error of Maximum Likelihood Estimates of Coefficients in Joint Models on Simulated Biomarker Data with both response and covariate missingness (The RMSEs are given in the brackets)}
	\begin{center}
	\begin{tabular}{|l|l|l|l|l|l|}
		\hline
		Variable  & \begin{tabular}[c]{@{}l@{}}Full \\ Dataset\end{tabular} & \begin{tabular}[c]{@{}l@{}}EMB\\ Algorithm\end{tabular} & \begin{tabular}[c]{@{}l@{}}PMM\\ Method\end{tabular} & \begin{tabular}[c]{@{}l@{}}NORM\\ Method\end{tabular} & \begin{tabular}[c]{@{}l@{}}CART \\ Method\end{tabular} \\
		\hline
		$\beta_1$   & 0.98  (0.12)                                                 & 
		0.85   (0.20)                                                & 
		0.90    (0.17)                                            & 
		0.83   (0.23)                                             & 
		0.80    (0.25)                                              \\
		$\beta_2$   & 0.96   (0.25)                                                & 
		0.82   (0.30)                                                & 
		0.87   (0.28)                                             & 
		0.80   (0.31)                                              & 
		0.78    (0.30)                                              \\
		$\beta_3$   & 0.96   (0.15)                                   & 
		0.82        (0.22)                                     & 
		0.80       (0.20)                                         & 
		0.78        (0.26)                                         & 
		0.76         (0.29)                                         \\
		$\beta_4$   & 0.95   (0.21)                        &
		 0.82        (0.29)                                      & 
		 0.85        (0.26)                                        & 
		 0.78         (0.31)                                        & 
		 0.76         (0.31)                                        \\
		$\beta_5$   & 0.97    (0.20)                       &
		 0.82         (0.29)                          & 
		 0.86         (0.23)                                  & 
		 0.80         (0.30)                                        & 
		 0.76         (0.30)                                         \\
		$\beta_6$   & 0.99    (0.13)                    &
		 0.80      (0.17)                                   & 
		 0.86      (0.16)                                        & 
		 0.81      (0.21)                                          & 
		 0.80       (0.25)                                          \\
		\hline
		$\gamma_1$  & 0.97   (0.12)                            &
		 0.83    (0.17)                                         & 
		 0.89     (0.15)                                          & 
		 0.81     (0.20)                                           & 
		 0.75      (0.24)                                           \\
		$\gamma_2$  & 0.96   (0.13)                                  &
		 0.84     (0.20)                                             & 
		 0.87     (0.18)                                          & 
		 0.81      (0.24)                                          & 
		 0.77       (0.30)                                          \\
		$\gamma_3$  & 0.98    (0.18)                      &
		 0.83  (0.22)                                                & 
		 0.87  (0.20)                                             & 
		 0.80   (0.26)                                             & 
		 0.79    (0.28)                                             \\
		\hline
		$\alpha$     & 0.98    (0.14)                       &
		 0.82        (0.19)                                           & 
		 0.88       (0.17)                                       & 
		 0.80         (0.24)                                        & 
		 0.76         (0.27)                                         \\
		D         & 0.97     (0.13)                          &
		 0.79     (0.20)                                              & 
		 0.83      (0.17)                                          & 
		 0.74      (0.24)                                           & 
		 0.73       (0.23)                                           \\
		$\sigma$ &      0.97 (0.06)                               &
		 0.81 (0.11)                                            &
		0.86 (0.09)                                            &
		0.77 (0.12)                                               &
		0.72 (0.15)\\ 
		\hline
	\end{tabular}
	\end{center}
\label{table:simcov_ml_data}
\end{table}

\begin{table}[!h]
	\caption{Coverage Probability and Root Mean Squared Error of Bayesian Estimates of Coefficients in Joint Models on Simulated Biomarker Data with only response missingness (The RMSEs are given in the brackets)}
	\begin{center}
	\begin{tabular}{|l|l|l|l|l|l|}
		\hline
		Variable & \begin{tabular}[c]{@{}l@{}}Full \\ Dataset\end{tabular} & \begin{tabular}[c]{@{}l@{}}EMB\\ Algorithm\end{tabular} & \begin{tabular}[c]{@{}l@{}}PMM\\ Method\end{tabular} & \begin{tabular}[c]{@{}l@{}}NORM\\ Method\end{tabular} & \begin{tabular}[c]{@{}l@{}}CART \\ Method\end{tabular} \\
		\hline
		$\beta_1$  & 0.97  (0.13)                           & 
		0.90       (0.20)                                            & 
		0.92        (0.17)                                        & 
		0.87        (0.22)                                         & 
		0.81         (0.26)                                         \\
		$\beta_2$  & 0.98   (0.21)                            &
		0.81     (0.26)                                        & 
		0.87      (0.24)                                          & 
		0.83      (0.28)                                           & 
		0.78       (0.29)                                           \\
		$\beta_3$  & 0.95   (0.13)                              &
		0.85     (0.23)                                              & 
		0.91      (0.221)                                          & 
		0.81      (0.25)                                           & 
		0.75       (0.31)                                           \\
		$\beta_4$  & 0.96  (0.17)                            & 
		0.82       (0.21)                                            & 
		0.87       (0.17)                                         & 
		0.80       (0.25)                                          & 
		0.77       (0.25)                                          \\
		$\beta_5$  & 0.97  (0.22)                           & 
		0.81      (0.23)                                             & 
		0.84      (0.25)                                          & 
		0.80      (0.28)                                          & 
		0.75       (0.31)                                           \\
		$\beta_6$  & 0.98  (0.11)                         & 
		0.83      (0.13)                                            & 
		0.86       (0.12)                                        & 
		0.80        (0.17)                                        & 
		0.73         (0.20)                                        \\
		\hline
		$\gamma_1$ & 0.97   (0.22)                          & 
		0.82   (0.26)                                               & 
		0.84    (0.27)                                           & 
		0.80     (0.30)                                           & 
		0.74      (0.31)                                          \\
		$\gamma_2$ & 0.98  (0.24)                         & 
		0.78    (0.29)                                              & 
		0.85   (0.26)                                           & 
		0.80     (0.30)                                           & 
		0.76     (0.32)                                            \\
		$\gamma_3$ & 0.96 (0.23)                           & 
		0.78  (0.28)                                                & 
		0.84   (0.26)                                            & 
		0.77    (0.31)                                            & 
		0.82    (0.35)                                             \\
		\hline
		$\alpha$    & 0.98  (0.15)                         & 
		0.86    (0.18)                                               & 
		0.90     (0.16)                                           & 
		0.82     (0.20)                                            & 
		0.84      (0.25)                                            \\
		D        & 0.98   (0.06)                          & 
		0.84      (0.15)                                             & 
		0.87       (0.10)                                         & 
		0.82        (0.19)                                         & 
		0.80        (0.22)                                          \\  
		$\sigma$      & 0.96 (0.11)            &
		0.84 (0.14)                                                 & 
		0.88 (0.14)                                             & 
		0.82 (0.17)                         & 
		0.77 (0.19)                        \\  
		\hline
	\end{tabular}
\end{center}
	\label{table:simwb_bayes_data}
\end{table}

\begin{table}[!h]
	\caption{Coverage Probability and Root Mean Squared Error of Maximum Likelihood Estimates of Coefficients in Joint Models on Simulated Biomarker Data with only response missingness (The RMSEs are given in the brackets)}
	\begin{center}
			\begin{tabular}{|l|l|l|l|l|l|}
				\hline
				parameters & Full Data & \begin{tabular}[c]{@{}l@{}}EMB\\ Algorithm\end{tabular} & \begin{tabular}[c]{@{}l@{}}PMM\\ Method\end{tabular} & \begin{tabular}[c]{@{}l@{}}NORM\\ Method\end{tabular} & \begin{tabular}[c]{@{}l@{}}CART \\ Method\end{tabular} \\
				\hline
				$\beta_1$    & 0.98 (0.12)     						& 
				0.90   (0.17)                                                & 
				0.93   (0.15)                                             & 
				0.88    (0.23)                                             & 
				0.83     (0.25)                                             \\
				$\beta_2$    & 0.96 (0.25)                          &
				0.89  (0.31)                                                 & 
				0.90    (0.28)                                            & 
				0.87    (0.30)                                             & 
				0.86    (0.30)                                             \\
				$\beta_3$    & 0.96 (0.15)                        &
				0.87    (0.20)                                               &
				0.89      (0.18)                                          & 
				0.81      (0.23)                                           &
				0.80       (0.25)                                           \\
				$\beta_4$    & 0.95 (0.21)                                 & 
				0.85        (0.27)                                           &
				0.88          (0.23)                                      & 
				0.83         (0.29)                                        & 
				0.81         (0.30)                                        \\
				$\beta_5$    & 0.97 (0.20)   								 &
				0.88    (0.27)                                               & 
				0.91    (0.21)                                            & 
				0.82     (0.28)                                            & 
				0.80     (0.29)                                             \\
				$\beta_6$    & 0.99 (0.13)                      			&
				0.86   (0.16)                                               &
				0.90    (0.14)                                           & 
				0.83    (0.20)                                           & 
				0.81    (0.23)                                             \\
				\hline
				$\gamma_1$   & 0.97 (0.12)                          &
				0.85      (0.17)                                             &
				0.91        (0.14)                                     & 
				0.87        (0.19)                                        &
				0.82         (0.22)                                        \\
				$\gamma_2$   & 0.96 (0.13)   & 
				0.87        (0.18)                                         &
				0.90        (0.17)                                      & 
				0.84        (0.22)                                        & 
				0.82         (0.26)                                        \\
				$\gamma_3$   & 0.98 (0.18)   							&
				0.86     (0.21)                                              &
				0.89     (0.20)                                          & 
				0.83      (0.25)                                           & 
				0.81      (0.26)                                           \\
				\hline
				$\alpha$      & 0.98 (0.14)    & 
				0.90      (0.16)                                       & 
				0.93       (0.16)                                        &
				0.88        (0.19)                                        & 
				0.84        (0.22)                                          \\
				$D$          & 0.97 (0.13)    & 
				0.86      (0.19)                                             & 
				0.89       (0.14)                                         & 
				0.82       (0.23)                                          & 
				0.80        (0.23)                                          \\
				$\sigma$          & 0.97 (0.06)    & 
				0.88      (0.08)                                             & 
				0.92       (0.08)                                         & 
				0.85       (0.11)                                          & 
				0.84        (0.13)                                          \\
				\hline
			\end{tabular}
	\end{center}
	\label{table:simwb_ml_data}
\end{table}

\begin{table}[!h]
	\caption{Comparison of Maximum Likelihood estimates of Coefficients in Joint Models on differently imputed Biomarker Data (The standard errors of the estimates are given in the brackets)}
	\begin{tabular}{|l|l|l|l|l|l|}
		\hline
		variables & \begin{tabular}[c]{@{}l@{}}Complete Data\\ (Excluding the \\ missing \\ Observations)\end{tabular} & \begin{tabular}[c]{@{}l@{}}EMB\\ Algorithm\end{tabular} & \begin{tabular}[c]{@{}l@{}}PMM\\ Method\end{tabular} & \begin{tabular}[c]{@{}l@{}}CART\\ Method\end{tabular} & \begin{tabular}[c]{@{}l@{}}NORM \\ Method\end{tabular} \\
		\hline
		(Intercept)                                                   & 
		4.232  (0.131)                                           & 
		4.283   (0.126)                                                & 
		4.222   (0.124)                                             & 
		4.212   (0.134)                                               & 
		4.258    (0.135)                                              \\
		(days/365)                                                    & 
		0.449  (0.139)                                         & 
		0.502  (0.088)                                                & 
		0.499  (0.094)                                              & 
		0.488  (0.105)                                               & 
		0.447  (0.108)                                                \\
		t\_stage$_3$                                                & 
		0.136     (0.115)                                          & 
		-0.334    (0.119)                                              & 
		0.041     (0.119)                                           & 
		-0.004    (0.125)                                           & 
		0.011      (0.126)                                            \\
		t\_stage$_4$                                                 & 
		-0.066    (0.124)                                           & 
		-0.119    (0.125)                                              & 
		-0.119    (0.125)                                           & 
		-0.142    (0.133)                                           & 
		-0.128    (0.135)                                             \\
		n\_stage$_1$                                                 & 
		0.150    (0.144)                                          & 
		0.061    (0.116)                                               & 
		0.161    (0.115)                                            & 
		0.184     (0.119)                                            & 
		0.137     (0.123)                                             \\
		n\_stage$_2$                                                 & 
		-0.128    (0.101)                                          & 
		-0.140    (0.086)                                              & 
		-0.116    (0.085)                                          & 
		-0.073    (0.089)                                           & 
		-0.148    (0.090)                                            \\
		trg\_score$_2$                                               & 
		0.067   (0.096)                                           & 
		0.110   (0.077)                                                & 
		0.122   (0.078)                                             & 
		0.112    (0.083)                                             & 
		0.118    (0.082)                                             \\
		trg\_score$_3$                                               & 
		0.047   (0.152)                                           & 
		0.135   (0.096)                                                & 
		0.135   (0.098)                                             & 
		0.115   (0.104)                                              & 
		0.109   (0.107)                                              \\
		trg\_score$_4$                                               & 
		0.356    (0.163)                                           & 
		0.149    (0.123)                                               & 
		-0.293   (0.122)                                           & 
		0.240    (0.127)                                             & 
		0.243    (0.129)                                              \\
		trg\_score$_5$                                               & 
		0.488      (0.218)                                         & 
		0.496      (0.225)                                             & 
		0.504      (0.240)                                         & 
		0.564      (0.234)                                           & 
		0.561      (0.242)                                            \\
		\hline
		t\_stage$_3$                                                   & 
		0.733    (1.041)                                           & 
		0.395    (0.974)                                               & 
		0.256    (0.962)                                            & 
		0.390    (0.967)                                            & 
		0.325    (0.972)                                              \\
		t\_stage$_4$                                                   & 
		1.991      (1.041)                                         & 
		1.439      (0.985)                                             & 
		1.315      (0.955)                                          & 
		1.423      (0.964)                                           & 
		1.364      (0.973)                                            \\
		n\_stage$_1$                                                   & 
		1.991     (0.933)                                          & 
		1.126     (0.800)                                              & 
		0.987      (0.807)                                          & 
		0.989      (0.807)                                           & 
		0.129      (0.816)                                           \\
		n\_stage$_2$                                                   & 
		0.784       (0.746)                                     & 
		0.684       (0.605)                                            & 
		0.641       (0.598)                                         & 
		0.582       (0.605)                                          & 
		0.697       (0.608)                                           \\
		\hline
		\begin{tabular}[c]{@{}l@{}}Association \\ Parameter ($\alpha$)\end{tabular} & 
		1.334         (0.240)                                    & 
		2.015         (0.221)                                          & 
		1.651         (0.214)                                       & 
		1.646         (0.216)                                        & 
		1.679         (0.219)                                         \\
		\begin{tabular}[c]{@{}l@{}}Variance of \\ random component\\ (D)\end{tabular} & 
		0.193  (0.214)                                            & 
		0.208  (0.251)                                                 & 
		0.188  (0.247)                                            & 
		0.565  (0.248)                                               & 
		0.600  (0.263)                                                \\
		AIC                                                         & 188.441                                                                                       & 415.523                                                 & 419.878                                              & 413.261                                               & 466.682                                               \\
		\hline
	\end{tabular}
	\label{table:real_ml_estimate}
\end{table}

\begin{table}[!h]
	\caption{Comparison of Bayesian estimates of Coefficients in Joint Models on differently imputed Biomarker Data (The posterior standard deviations of the estimates are given in the brackets)}
	\begin{tabular}{|l|l|l|l|l|l|}
		\hline
		Variable                                                                & \begin{tabular}[c]{@{}l@{}}Complete Data\\ (Excluding the \\ missing \\ Observations)\end{tabular} & \begin{tabular}[c]{@{}l@{}}EMB\\ Algorithm\end{tabular} & \begin{tabular}[c]{@{}l@{}}PMM\\ Method\end{tabular} & \begin{tabular}[c]{@{}l@{}}CART\\ Method\end{tabular} & \begin{tabular}[c]{@{}l@{}}NORM \\ Method\end{tabular} \\
		\hline
		(Intercept)                                                      & 
		4.211     (0.428)                                              & 
		4.273     (0.293)                                             & 
		4.227     (0.312)                                           & 
		4.218     (0.287)                                            & 
		4.242     (0.314)                                             \\
		(days/365)                                                                & 
		0.536       (0.188)                                          & 
		0.548       (0.154)                                         & 
		0.562       (0.145)                                         & 
		0.536       (0.148)                                          & 
		0.529       (0.165)                                           \\
		t\_stage$_3$                                                             & 
		0.136      (0.410)                                            & 
		0.006       (0.297)                                           & 
		0.041       (0.307)                                         & 
		-0.002      (0.285)                                          & 
		0.019       (0.310)                                           \\
		t\_stage$_4$                                                             & 
		-0.033        (0.432)                                          & 
		-0.083         (0.298)                                         & 
		-0.108         (0.316)                                      & 
		-0.115         (0.291)                                       & 
		-0.102          (0.314)                                       \\
		n\_stage$_1$                                                             & 
		0.176          (0.449)                                       & 
		0.061          (0.283)                                     & 
		0.159          (0.293)                                      & 
		0.183          (0.266)                                       & 
		0.146          (0.285)                                        \\
		n\_stage$_2$                                                             & 
		-0.121         (0.313)                                          & 
		-0.143         (0.209)                                        & 
		-0.142         (0.213)                                      & 
		-0.079         (0.198)                                       & 
		-0.144         (0.214)                                        \\
		trg\_score$_2$                                                           & 
		0.071        (0.323)                                            & 
		0.096         (0.131)                                       & 
		0.125         (0.155)                                       & 
		0.102         (0.100)                                       & 
		0.115         (0.151)                                         \\
		trg\_score$_3$                                                           & 
		0.045         (0.521)                                           & 
		0.107         (0.163)                                        & 
		0.149         (0.152)                                      & 
		0.092         (0.168)                                        & 
		0.114         (0.164)                                         \\
		trg\_score$_4$                                                           & 
		0.345         (0.550)                                           & 
		0.171         (0.239)                                        & 
		0.232         (0.176)                                       & 
		0.186         (0.168)                                        & 
		0.216         (0.185)                                         \\
		trg\_score$_5$                                                           & 
		0.433         (0.793)                                         & 
		0.482         (0.546)                                         & 
		0.565         (0.626)                                       & 
		0.545         (0.579)                                        & 
		0.544         (0.602)                                         \\
		\hline
		t\_stage$_3$                                                               & 
		1.312      (1.290)                                              & 
		0.513      (1.091)                                            & 
		0.508       (1.037)                                         & 
		0.461       (1.085)                                          & 
		0.563       (1.103)                                          \\
		t\_stage$_4$                                                               & 
		2.739      (1.295)                                               & 
		1.553      (1.133)                                           & 
		1.567      (1.083)                                          & 
		1.503      (1.066)                                           & 
		1.631      (1.117)                                           \\
		n\_stage$_1$                                                               & 
		2.482       (1.106)                                             & 
		1.195       (0.850)                                        & 
		1.080       (0.841)                                         & 
		1.094       (0.856)                                         & 
		1.167       (0.848)                                           \\
		n\_stage$_2$                                                  & 
		1.156        (0.895)                                            & 
		0.709        (0.643)                                          & 
		0.669        (0.642)                                        & 
		0.630        (0.648)                                         & 
		0.690        (0.630)                                          \\
		\hline
		\begin{tabular}[c]{@{}l@{}}Association \\ Parameter ($\alpha$)\end{tabular}             & 1.842      (0.687)                                           & 
		1.916      (0.570)                                           & 
		1.163      (0.655)                                          & 
		1.078      (0.515)                                           & 
		1.156      (0.441)                                            \\
		\begin{tabular}[c]{@{}l@{}}Variance of \\ random component \\ (D)\end{tabular} & 
		0.728    (0.215)                                          & 
		0.596    (0.246)                                               & 
		0.599    (0.248)                                            & 
		0.565     (0.241)                                            & 
		0.600    (0.272)                                              \\
		DIC                                                                     & 341.982                                                                                         & 665.964                                                  & 667.765                                              & 646.927                                               & 730.203                                               \\
		\hline
	\end{tabular}
\label{table:real_bayes_estimate}
\end{table}

\end{document}